\documentstyle[12pt]{article}
\newcommand{\be}{\begin{equation}}
\newcommand{\ee}{\end{equation}}
\newcommand{\bea}{\begin{eqnarray}}
\newcommand{\eea}{\end{eqnarray}}
\newcommand{\nn}{\nonumber}

\begin{document}

\begin{titlepage}
\mbox{ }
\rightline{BARI-TH/95-191}
\rightline{UCT-TP-222/95}
\rightline{January 1995}
\vspace{1.0cm}
\begin{center}
\begin{Large}
{\bf  Tests of factorization in color suppressed\\[.5cm]
	nonleptonic B decay modes}
\end{Large}

\vspace{0.8cm}

{\large \bf P. Colangelo$^{a}$,
C.A. Dominguez$^{b,}$\footnote{John Simon Guggenheim Fellow 1994-1995},
N. Paver$^{c,}$\footnote{Also supported by MURST, Ministero dell'Universit\'a
e Ricerca Scientifica e Tecnologica, Italy}
}\\
\end{center}

\vspace{0.2cm}

\begin{center}
\small
\begin{itemize}
\item[$^{a}$]Istituto Nazionale di Fisica Nucleare, Sezione di Bari, Italy
\item[$^{b}$]Institute of Theoretical Physics and Astrophysics,
	University of Cape Town, Rondebosch 7700, South Africa
\item[$^{c}$]Dipartimento di Fisica Teorica, Universit\`{a} di Trieste,
 Italy, and Istituto Nazionale di Fisica Nucleare, Sezione di Trieste, Italy
\end{itemize}
\end{center}

\vspace{0.5cm}

\begin{abstract}
\noindent
We propose a number of tests of factorization for the color-suppressed
nonleptonic decay channels: $B \to K^{(*)} \; \eta_c$ and
$B \to K^{(*)} \; \eta^\prime_c$.
The relevant leptonic constants and form factors are calculated within the
QCD sum rules approach.
\end{abstract}
\end{titlepage}

\setcounter{footnote}{0}
\setcounter{page}{1}
\setlength{\baselineskip}{1.5\baselineskip}

\noindent
\subsection*{1. Introduction}
It has been recently observed \cite{Gourdin, Aleksan} that the available
experimental data \cite{Stone} on the longitudinal polarization of the final
particles in the decay $B \to K^* \; J/\Psi$:
$\rho_L=\Gamma(B \to K^* \; J/\Psi)_{LL}/\Gamma(B \to K^* \; J/\Psi)
=0.84 \pm 0.06\pm 0.08$,
together with the ratio
$R_{J/\Psi}=\Gamma(B \to K^* \; J/\Psi)/\Gamma(B \to K \; J/\Psi)
=1.64 \pm 0.34$, can
severely constrain the semileptonic $B \to K^{(*)}$ form factors
if  these nonleptonic transitions are evaluated within the factorization
approach.\par
Neglecting penguin operators, the relevant effective Hamiltonian density is
\be
{\cal H}_W = - {G_F \over \sqrt 2} \;
V^*_{c s} V_{c b} \; ( c_1(\mu) \; O_1 + c_2(\mu) \; O_2)\; +\; h.c.
\hskip 2pt , \label{ham} \ee
where
$O_1= (\bar s_i c_i)_{V-A} (\bar c_j b_j)_{V-A}$ and
$O_2= (\bar s_i c_j)_{V-A} (\bar c_j b_i)_{V-A}$
($i,j$ are color indices, $V-A$ refers to $ \gamma_\mu ( 1- \gamma_5) $,
$G_F$ is the Fermi constant and $V_{lm}$ are Cabibbo-Kobayashi-Maskawa matrix
elements). The values of the Wilson coefficients at the next-to-leading order
in QCD can be taken as $c_1(m_b)=1.08\pm0.01$ and $c_2(m_b)=-0.19\pm0.02$
\cite{Buras,Buras1}.\par
The above transitions are color-suppressed. After Fierz reordering
Eq. (\ref{ham}), in the framework of factorization and vacuum saturation
dominance, the factorized amplitude is
\be
A (B \to K^{(*)} J/\Psi) =
- {G_F \over \sqrt 2} V^*_{cs} V_{cb}
 a_2   < K^{(*)} | \bar s \gamma_\mu (1- \gamma_5) b | B>
\;<J/\Psi |\bar c \gamma^\mu (1- \gamma_5) c |0> \hskip 2pt ,
\label{amp}\ee
where $a_2=c_2+c_1/N_c$, with $N_c$ is the number of colors. Eq. (\ref{amp})
involves the matrix element
\be
<J/\Psi (q, \; \epsilon ) | \bar c \gamma_\mu c | 0>   =
 f_{J/\Psi}  \; M_{J/\Psi} \; \epsilon^*_\mu\hskip 2pt ,\label{leptonic}\ee
where the leptonic constant $f_{J/\Psi}$ is directly obtained from the measured
decay $J/\Psi\to \ell^+ \ell^-$.
Equation (\ref{amp}) also involves the semileptonic $B  \to K^{(*)}$ matrix
elements ($q = p - p^\prime$):\footnote{We adopt
the Bauer-Stech-Wirbel parameterization \cite{bsw} where:
$F_0(0)=F_1(0)$, $A_3(0)=A_0(0)$ and
\\  \noindent
$A_3 (q^2)\;=\;{[(M_{B}+M_{K^*}) / 2 M_{K^*}]} A_1 (q^2) -
{[(M_{B}-M_{K^*}) / 2 M_{K^*}]} A_2 (q^2)$.}
\be
< K (p^\prime) | \bar s \gamma_\mu b | B(p)> =
 (p + p^\prime)_\mu F_1 (q^2) +
{ M^2_B - M^2_K \over q^2 }
[F_0(q^2) - F_1(q^2)] q_\mu\hskip 2pt ,\label{spinzero}\ee
\noindent and
\newpage
\bea
 <K^* (p^\prime, \eta) | \bar s \gamma_\mu (1- \gamma_5) b
| B(p)>  &=&
 \eta_{\mu \nu \rho \sigma}
\eta^{* \nu} p^\rho p^{\prime \sigma} { 2 V(q^2) \over M_B + M_{K^*}}
 -  i \Big[ (M_{B} + M_{K^*}) A_1 (q^2) \eta_{\mu}^{\ast} \; -
\nn \\
 -   {A_2 (q^2)
\over M_{B} + M_{K^*}} (\eta^{\ast}\cdot p) (p + p^\prime)_\mu
&-&  \;(\eta^{\ast}\cdot p) {2 M_{K^*} \over q^2} q_\mu
\big(A_3 (q^2) - A_0 (q^2)\big) \Big] \hskip 2pt .
\label{eqi1} \eea
Actually, in Eq. (\ref{amp}) the form factors must be evaluated at
$q^2=M^2_{J/\Psi}$ and, in particular, the quantities $\rho_L$ and
$R_{J/\Psi}$ depend on the ratios:
$V(M^2_{J/\Psi}) / A_1(M^2_{J/\Psi})$,
$A_2(M^2_{J/\Psi}) / A_1(M^2_{J/\Psi})$
and
$F_1(M^2_{J/\Psi}) / A_1(M^2_{J/\Psi})$.\par
Using predictions of popular theoretical models for the form
factors,\footnote{For a rather complete set of references
on heavy-to-light meson
semileptonic form factors see, e.g., Refs. \cite{Gourdin,Gourd,Gourdin1}.}
difficulties are met in simultaneously reproducing the experimental outcomes
for $\rho_L$ and $R_{J/\Psi}$ \cite{Gourdin, Aleksan}. This has prompted a
number of interesting phenomenological analyses trying to elucidate the
source of the problem. On the one side, the blame could be put on
factorization which, in principle, can receive substantial corrections
from the nonfactorizable component of Eq. (\ref{ham}), that
has the large coefficient $c_1$ \cite{kamal}. At present, such effects are
quite difficult to predict by a dynamical calculation.\par
On the other side, the alternative attitude would be to accept factorization,
which is an appealing exact feature of multicolor chromodynamics in the
limit $N_c \to \infty$ \cite{Thooft},\footnote{Arguments in support of
factorization for finite $N_c$, in the limit of infinitely heavy quarks,
can be found in \cite{Dugan} (with some criticisms in \cite{Aglietti}), and
different analyses of the role of nonfactorizable matrix elements
for selected $B$ and $D$  nonleptonic decays \cite{Shifman, Khod}
do not seem to indicate sizable deviations if $1/N_c$ corrections are
collectively discarded \cite{BWS1}.} and focus, instead, on form factor
predictions. Actually, not all form factor models are quite predictive in the
region of large $q^2\simeq M^2_{J/\Psi}$, and in these cases the
$q^2$-dependence has to be assumed, e.g., by invoking lowest-lying meson
dominance and/or quark counting rules. In this context, assuming factorization,
different phenomenological $q^2$-dependences have been scrutinized in
Ref. \cite{Gourd}, attempting to reproduce the available data.\par
Clearly, experimental studies of other color-suppressed $B$ decays should be
very desirable in order to complete the present information and more
extensively test the factorization scheme. From this point of view, an
interesting role is played by the transitions $B \to K^{(*)}\; \eta_c $,
that differ from the previous ones,  $B \to K^* \; J/\Psi$, just by the
charm-quark spin configuration. These processes, not measured yet,
have been
considered also in the phenomenological analyses of
Refs. \cite{Gourd,Gourdin1}, where relations between the branching ratios of
$B \to K^{(*)}\; \eta_c$ and $B \to K^{(*)}\; J/\Psi$ in terms of ratios of
leptonic constants and form factors have been worked out.\par
In this note, we consider predictions for $B \to K^{(*)}\; \eta_c $ and
$B \to K^{(*)}\; \eta^\prime_c$ that can be obtained in the factorization
approximation using the method of QCD sum rules. The aim of our discussion
differs from \cite{Gourd,Gourdin1} in that we adopt here a specific
theoretical framework to estimate consistently from the outset all
nonperturbative quantities necessary to predict these processes. Indeed, in
this case the relevant leptonic constants in Eq. (\ref{amp}),
where $J/\Psi$ must be replaced by $\eta_c$ or $\eta^\prime_c$, are
\begin{equation}<\eta_c\; (\eta^\prime_c)| \bar c \gamma_\mu\gamma_5 c |0> =
  - i f_{\eta_c\; (\eta^\prime_c)}\; q_\mu\; ,\label{effetac}\end{equation}
that are not experimentally known, but must be determined theoretically.
Previous references rely on constituent quark model arguments to relate
$f_{\eta_c}$ to $f_{J/\Psi}$. In fact, leptonic constants represent a natural
field of application for QCD sum rules \cite{RRY}. The virtue of this
approach, being fully relativistic and field-theoretic by construction,
is that it allows to avoid the notion of confined-quark wavefunction,
and it incorporates
naturally fundamental features of QCD, such as perturbative asymptotic
freedom and nonperturbative quark and gluon condensation. Thus, as a
computational scheme alternative to the constituent quark model, it should be
sensible to present an updated estimate of $f_{\eta_c}$ and
$f_{\eta^\prime_c}$ in the framework of QCD sum rules. We recall that
besides their relevance to the $B$ decays under consideration,
emphasized in Ref. \cite{Desh}, these constants are of phenomenological
interest also to other applications of QCD sum rules, such as the radiative
processes $J/\Psi \to \eta_c\; \gamma$ and
$\eta_c\;(\eta^\prime_c)\to\gamma\gamma$ \cite{Rad}.\par
The other significant aspect of the present calculation is that the
needed $B\to K^{(*)}$ form factors $F_0(q^2)$ and $A_0(q^2)$ (see
Eqs. (\ref{spinzero}) and (\ref{eqi1})) have not been studied as extensively
as others. Since these
form factors must be evaluated at $q^2=M_{\eta_c}^2$ or
$q^2=M_{\eta^\prime_{\hskip 2pt c} }^2$, we can exploit the capability of QCD
sum rules to provide an evaluation of their $q^2$-dependence, within
some theoretical uncertainty. This would complete the phenomenological
description of color-suppressed $B$ decays to charmonium based on QCD sum
rules and factorization, by the addition of a set of predictions
which, ultimately, could be interesting to compare to experimental data (when
available) or to predictions from alternative nonperturbative schemes.\par
The paper is organized as follows. In Sec. 2 we present a QCD sum rule
calculation of the leptonic constants
$f_{\eta_c}$ and $f_{\eta^\prime_c}$,
together with a comparison with previous determinations. In Sec. 3
we briefly describe a calculation of the form factors
$F_0(q^2)$ and $A_0(q^2)$ for the transitions $B \to K, K^*$; finally,
in Sec. 4 we give predictions for ratios such as
$\Gamma(B \to K^* \eta^\prime_c)/\Gamma(B \to K^* \eta_c)$
and $\Gamma(B \to K \eta^\prime_c)/\Gamma(B \to K \eta_c)$.

\subsection*{2. QCD sum rule calculation of $f_{\eta_c}$ and
 $f_{\eta^\prime_c}$}
In order to determine the leptonic constants of the $\eta_{c}$
and $\eta'_{c}$ mesons from QCD sum rules we consider the two-point function
\bea
\psi_{5}(q) = i \int d^{4}x \; e^{iqx}\; < 0| T (\partial^{\mu} A_{\mu} (x)
	\partial^{\nu} A_{\nu}^{\dagger} (0)) |0 > \; ,
\eea
where $\partial ^{\mu} A_{\mu} (x) = 2 m_{c} : \bar{c}(x) i \gamma_{5}
c(x):$. This two-point function is known in perturbative QCD to two-loop
order \cite{BROAD}, and the leading non-perturbative term in the
Operator Product Expansion (OPE) is also known \cite{RRY} (it is given
in terms of the gluon condensate). We exploit two different types
of QCD sum rules, viz. Hilbert transforms at $Q^{2} = 0$, and Laplace
transforms. The former can be written as
\bea
M_{n}(0) \equiv \frac{1}{\pi} \; \int^{\infty}_{0} \;
\frac{ds}{s^{n+2}} \; \mbox{Im} \; \psi_{5}(s) \;.
\eea
The perturbative QCD contribution is given by \cite{BROAD}
\be
M_{n}(0)|_{PT} = \frac {3}{4 \pi^{2}} \; \frac
{\Gamma^{2}(n)}{\Gamma(2n)} \; \frac {1}{(2n+1)} \;
\left( \frac{1}{m_{c}^{2}} \right)^{n-1} (1 + a_{n} \alpha_{s})\; ,
\ee
where
\begin{eqnarray*}
a_{n} = \frac{4}{3 \pi} \left\{ 1 - \frac{1}{2n} - \frac{3}{n+1} -
     \frac{3}{2(n+2)} + \left( \frac{1}{n} + \frac {1}{n+1} \right)
     \right.
\end{eqnarray*}
\begin{eqnarray}
\times \sum_{r=1}^{n+1} \left[ \frac {n+2}{r} \right.
\left. \left. \frac{B(r, \frac{1}{2})}{B(n, \frac{1}{2})} - \frac{3}{r} +
\frac{2}{2r-1} \right] \right\} \; ,
\end{eqnarray}

and $B(x,y)$ is the beta function. The leading non-perturbative term in
the OPE gives the following contribution \cite{RRY}
\be
M_{n}(0)|_{NP} = - \frac {1}{24} <\frac{\alpha_{s}}{\pi} G^{2}>
\frac{1}{(4m_{c}^{2})^{n+1}}
 \frac{(n-3) \Gamma(\frac{1}{2}) \Gamma(n+3)}
     {\Gamma(n+ 5/2)} \; .
\ee
The hadronic spectral function is parametrized by the standard ansatz
\bea
\frac{1}{\pi} \; \mbox{Im} \; \psi_{5} (s)|_{HAD} = f_{\eta_{c}}^{2}
M_{\eta_{c}}^{4} \; \delta \left( s - M_{\eta_{c}}^{2} \right)
+ \theta (s - s_{0})  \; \frac{1}{\pi} \; \mbox{Im} \;
\psi_{5} (s)|_{PT} \; ,
\label{ans}
\eea

where a second pole term for the $\eta'_{c}$ may be added to Eq. (\ref{ans}).
Putting everything together, and ignoring as a
first step the contribution of the $\eta'_{c}$ meson,
the Hilbert transform QCD sum rules become
\begin{eqnarray*}
\frac{f_{\eta_{c}}^{2}}{M_{\eta_{c}}^{2n}} =
\int_{4m_{c}^{2}}^{s_0} \; \frac{ds}{s^{n+2}} \; \frac{1}{\pi} \;
\mbox{Im} \; \psi_{5}(s)|_{PT}
 - \frac {1}{24} \; < \frac{\alpha_{s}}{\pi} G^{2} > \;
\frac{1}{(4m_{c}^{2})^{n+4}} \;
\end{eqnarray*}
\begin{eqnarray}
\times \frac{(n-3) \Gamma(\frac{1}{2}) \Gamma(n+3)}{\Gamma(n+5/2)} \; .
\end{eqnarray}

We take ratios of two consecutive moments for various values of $n$
in order to find the duality window (range of values of $s_{0}$) for
which the mass of the $\eta_{c}$ is correctly predicted. Once this mass
is accounted for, the leptonic decay constant follows from any given
moment. In addition to expecting some region of stability in
$s_{0}$, one also expects the mass and decay constant to be reasonably
independent of $n$. These expectations are partly fulfilled by the results
of our calculation, as may be appreciated from Figs. 1 and 2 where we show
$M_{\eta_{c}}$ and
${\tilde f}_{\eta_{c}}=f_{\eta_{c}}/\sqrt 2$
as a function of $n$ for the particular
choice: $\Lambda = 200 \; \mbox{MeV}$, $m_{c} = 1.39 \; \mbox{GeV}$,
and $<\alpha_{s} G^{2}> = 0.063 \; \mbox{GeV}^{4}$. The solid (dash) curve
in Fig. 1 is the theoretical (experimental) result. In Fig. 2 the solid
(dash) curve is obtained by using the theoretical (experimental) value
of the $\eta_{c}$ mass in the Hilbert moments. While the stability in
$n$ is more than adequate, there is some sensitivity to changes in $s_{0}$.
This feature is well known from applications of QCD sum rules to heavy-light
quark systems. The situation here, in the case of two heavy quarks, is then
not much different. We have explored the following range of parameters,
dictated by the gluon condensate and quark-mass analyses of \cite{BERT} and
\cite{DGP}, respectively: $<\alpha_{s} G^{2}> = 0.063 - 0.19\;
\mbox{GeV}^{4}$, $m_{c} = 1.46 \pm 0.07 \; \mbox{GeV}$. The QCD scale was
varied in the range $\Lambda = 200 - 300 \; \mbox{MeV}$. The values of $s_{0}$
needed to correctly reproduce the experimental value of $M_{\eta_{c}}$
were found in the range $s_{0} = 9.5 - 10.5 \; \mbox{GeV}^{2}$. Results for
$M_{\eta_{c}}$ and $f_{\eta_{c}}$ corresponding to various choices of
$\Lambda$, $m_{c}$, the gluon condensate and $s_{0}$ are qualitatively
similar to Figs. 1 and 2. Quantitatively, we find
$f_{\eta_{c}} \simeq 256 - 300 \; \mbox{MeV}$.

In a second step, we include a second pole in the spectral function
Eq. (\ref{ans}) to account for the $\eta'_{c}$ meson, and repeat the analysis.
We find that, in order to reproduce correctly the experimental values of
both  $M_{\eta_{c}}$ and $M_{\eta'_{c}}$
(for the mass of $\eta'_{c}$
we use the value
$M_{\eta'_{c}}=3595\pm5\;MeV$ \cite{Ed}),
the value of $s_{0}$ must
increase by some 30\% ($s_{0}\simeq 13 - 14\; \mbox{GeV}^{2}$).
This increase is to be expected, since now
$s_{0}\ge M_{\eta'_{c}}^{2}$.
 We find:
\be
f_{\eta_{c}} \simeq 301 - 326 \; \mbox{MeV},\;\;\;\;\;\;\;\;
f_{\eta'_{c}} \simeq 231 - 255 \; \mbox{MeV}\;
\ee
and, therefore, the result for
$f_{\eta_{c}}$  is now somewhat higher.
The reason why the uncertainty in $f_{\eta_{c}}$ is now smaller
than before is easy to understand. With a two-pole spectral
function, the sum rules are now constrained to reproduce
correctly the two meson masses simultaneously. This constraint effectively
reduces the size of the parameter space spanned by
$\Lambda$, $m_{c}$, $<\alpha_{s} G^{2}>$ and $s_{0}$.

Next, we make use of the Laplace transform QCD sum rule and its first
derivative \cite{BERT2}
\begin{eqnarray*}
 f_{\eta_{c}}^{2} \; M_{\eta_{c}}^{4} e^{-M_{\eta_{c}}^{2}/M_{L}^{2}}
\; = \; \int_{4m_{c}^{2}}^{s_{0}}  \; ds \; e^{-s/M_{L}^{2}}
\; \frac{1}{\pi} \; \mbox{Im} \; \psi_{5}(s)|_{PT}
\end{eqnarray*}
\begin{eqnarray}
- \frac{m_{c}^{2}}{6 \sqrt{\pi}} <\alpha_{s} G^{2}>
e^{-4 m_{c}^{2}/M_{L}^{2}}
\left[ G \left(- \frac{3}{2}, \frac{3}{2},
\frac{4m_{c}^{2}}{M_{L}^{2}} \right) -
6G \left( - \frac{1}{2}, \frac{3}{2}, \frac{4m_{c}^{2}}{M_{L}^{2}} \right)
\right]
\label{sr}
\end{eqnarray}

\begin{eqnarray*}
 f_{\eta_{c}}^{2} \; M_{\eta_{c}}^{6} e^{-M_{\eta_{c}}^{2}/M_{L}^{2}}
\; = \; \int_{4m_{c}^{2}}^{s_{0}}  \; ds \; s \; e^{-s/M_{L}^{2}}
\; \frac{1}{\pi} \; \mbox{Im} \; \psi_{5}(s)|_{PT}
\end{eqnarray*}
\begin{eqnarray*}
- 4m_{c}^{2} \; \frac{m_{c}^{2}}{6\sqrt{\pi}} \; <\alpha_{s} G^{2} > \;
e^{-4 m_{c}^{2}/M_{L}^{2}}
\left\{ G \left(- \frac{3}{2}, \frac{3}{2},
\frac{4m_{c}^{2}}{M_{L}^{2}} \right) \right.
\end{eqnarray*}
\begin{eqnarray}
\left. - 6G \left( - \frac{1}{2}, \frac{3}{2}, \frac{4m_{c}^{2}}{M_{L}^{2}}
\right)
- \frac{3}{2} G \left( - \frac{1}{2}, \frac{3}{2}, \frac{4m_{c}^{2}}
{M_{L}^{2}} \right) + 3 G \left( \frac{1}{2}, \frac{3}{2}, \frac{4m_{c}^{2}}
{M_{L}^{2}} \right) \right\}
\label{sr1}
\end{eqnarray}

where $M_{L}^{2}$ is the Laplace variable and the function $G(b,c,\omega)$
\be
G(b,c,\omega) = \frac{\omega^{-b}}{\Gamma(c)} \;
\int_{0}^{\infty} \; dt \; t^{c-1} \; e^{-t}
\left(1 + \frac{t}{\omega} \right)^{-b}
\ee

is related to the Whittaker function $W_{\lambda,\mu}(\omega)$ through
\cite{ABR}
\be
G(b,c,\omega) = \omega^{\mu - \frac{1}{2}} \;
e^{\omega/2} W_{\lambda}, \mu(\omega) \; .
\ee
Considering first a one-pole spectral function, the ratio of Eqs. (\ref{sr1})
and (\ref{sr}) determines $M_{\eta_{c}}$ as a function of the Laplace variable
$M_{L}^{2}$ and of $s_{0}$. We find that $M_{\eta_{c}}$ is
stable against changes in $M_{L}^{2}$ in the wide region
$M_{L}^{2} \simeq 1 - 5 \; \mbox{GeV}^{2}$. The experimental value of
$M_{\eta_{c}}$ is correctly reproduced for $s_{0} \simeq 10 \;
\mbox{GeV}^{2}$, roughly independent of the values of $\Lambda$, $m_{c}$
and $<\alpha_{s} G^{2}>$. This value of $s_{0}$ is in agreement with the
one obtained from the Hilbert moments. In Fig. 3 we show the result for
$M_{\eta_{c}}$ (solid line) compared to the experimental value (dash line),
for $\Lambda = 200 \; \mbox{MeV}$, $m_{c} = 1.39 \; \mbox{GeV}$ and
$<\alpha_{s} G^{2}> = 0.063 \;  \mbox{GeV}^{4}$. The result for
${\tilde f}_{\eta_{c}}=f_{\eta_{c}}/\sqrt 2$
obtained by using the predicted (experimental) value of
$M_{\eta_{c}}$ is shown in Fig. 4 as the solid (dash) curve. Exploring
the parameter space of $\Lambda$, $m_{c}$, $<\alpha_{s} G^{2}>$
and $s_{0}$ leads to the prediction:
$f_{\eta_{c}} \simeq 265 - 274 \; \mbox{MeV}$, in agreement
with the Hilbert moments.

Adding a second pole to the hadronic spectral function, considering
higher derivatives of Eq. (\ref{sr}) in $1/M^2_L$,
and redoing the analysis we find that
it is possible to reproduce the experimental values of both
$M_{\eta_{c}}$  and $M_{\eta'_{c}}$ for $s_{0}\simeq 13 - 14 \; \mbox{GeV}
^{2}$. At the same time,
the stability region in the Laplace variable $M_{L}^{2}$ remains
wide: $M_{L}^{2} \simeq 2 - 5 \; \mbox{GeV}^{2}$. The
predictions for the leptonic decay constants are
\be
f_{\eta_{c}} \simeq 292 - 310 \; \mbox{MeV},\;\;\;\;\;\;\;\;
f_{\eta'_{c}} \simeq 247 - 269 \; \mbox{MeV}\;,
\ee
in agreement with the results from the Hilbert moments, Eq. (14).
Combining the predictions from the two methods we obtain our final result
\be
f_{\eta_{c}} = 309 \pm  17 \; \mbox{MeV},\;\;\;\;\;\;\;\;
f_{\eta'_{c}} = 250 \pm 19 \; \mbox{MeV}\;,
\label{pred}
\ee
\be
\frac{f_{\eta'_{c}}} {f_{\eta_{c}}} = 0.8 \pm 0.1 \; .
\ee
Estimates of
$f_{\eta_{c}}, f_{\eta'_{c}}$, and of the $J/\Psi$ leptonic constant
were made many
years ago, in the early days of the QCD sum rules approach
 \cite{NOV,Rein,Shif1}. For example, in
\cite{NOV}  a QCD sum rule determination
of $f_{\eta_{c}}$
has been carried out at
a one-loop approximation in perturbative
QCD, and no gluon condensate corrections, in the framework of Hilbert
moments. With an input value of the charm
quark mass: $m_{c} = 1.25\; \mbox{GeV}$ (to be compared
with the recent determination \cite{DGP} employed here:
$m_{c} = 1.46 \pm 0.07\; \mbox{GeV}$ ), and the choice
$s_{0} = 16 \;\mbox{GeV}^{2}$, the value
$f_{\eta_{c}} = 252 - 369 \; \mbox{MeV}$ has been obtained.
Our determination here represents
an improvement in many respects: radiative (two-loop) as well as
non-perturbative corrections have been incorporated into the theoretical
side of the sum rules, Hilbert and Laplace sum rules have been employed,
updated values of the input parameters have been adopted, and the
analysis was constrained to reproduce correctly the masses of both
pseudoscalar mesons: $M_{\eta_{c}}$ and $M_{\eta'_{c}}$.

\subsection*{3. $F_0(q^2)$ and $A_0(q^2)$ from QCD sum rules}

The form factors
$F_0(q^2)$ and $A_0(q^2)$, related to the weak transitions
$B \to K$ and $B \to K^*$, can be computed by
three-point function QCD sum rules, following the same strategy adopted
in the calculation of the leptonic constants
$f_{\eta_c}$ and
$f_{\eta^\prime_c}$.
The starting point is provided by the correlators
\cite{ioffe, shiflibro} :
\be
\Pi_{\mu\nu} (p, p^\prime, q) = i^2 \int dx \; dy \;
e^{i ( p^\prime \cdot x - p \cdot y)} \;
<0| T \{ j^{K}_\nu(x) V_\mu(0) j^\dagger_5(y) \} |0> \;  \label{corrk}
\ee
\noindent and
\be
T_{\mu\nu} (p, p^\prime, q) = i^2 \int dx \; dy \;
e^{i ( p^\prime \cdot x - p \cdot y)} \;
<0| T \{ j^{K^*}_\nu(x) A_\mu(0) j^\dagger_5(y) \} |0> \; ; \label{corrks}
\ee
\noindent the quark currents
$j_5$ and $j_\nu^{K,K^*}$
are given by:
$j_5 = \bar d i \gamma_5 b$,
$j^{K}_\nu = \bar d \gamma_\nu \gamma_5 s$,
$j^{K^*}_\nu = \bar d \gamma_\nu  s$, whereas
the flavour changing weak currents
$V$ and $A$ are: $V_\mu =\bar s \gamma_\mu b$ and
$A_\mu =\bar s \gamma_\mu \gamma_5 b$.

The products:
$q^\mu \Pi_{\mu \nu}$ and $q^\mu T_{\mu \nu} $
receive a hadronic contribution from the  states $B, K$ and $B, K^*$,
respectively, so that they can be expressed in terms of the form factors
$F_0(q^2)$ and $A_0(q^2)$ and of
a continuum of states;
on the other hand, a QCD calculation can be performed for the same operator
products, in the limit of large and spacelike $p^2$ and $p^{\prime 2}$, in
terms of the perturbative QCD contribution and non-perturbative power
corrections, proportional to vacuum matrix elements of high dimensional
operators.
The matching of the hadronic and QCD representations of the correlators
can be improved by a double Borel transform in the variables $-p^2$ and
$-p^{\prime 2}$. We omit here the details of the calculation, and simply
present the result of such a procedure for the form factor $F_0(q^2)$.
Including power corrections in the OPE up to
dimension 5 condensates, one obtains the following equation:
\bea
f_K f_B {M^2_B \over m_b} (M^2_B - M^2_K) F_0(q^2) & = &
 {3 \over 8 \pi^2} \int_D ds \; ds^\prime
\rho (s, s^\prime, q^2) \;
 \exp [{M^2_B-s \over M^2} + {M^2_K-s^\prime \over M^{\prime 2}} ]
 \nonumber \\
  -  (m_b-m_s) \Big\{ m_b  \; <\bar q q>  &-&  d_5 \;
<\bar q g_s \sigma^{\mu \nu} G^a_{\mu \nu} {\lambda^a \over 2 } q>
\Big\}
\; \exp[ { M^2_B -m^2_b \over M^2}+{ M^2_K -m^2_s \over M^{'2}}]
 \nonumber \\
\label{f0} \eea
where the spectral function $\rho$ is given by:
\be
\rho(s,s',q^2)= {1 \over \sqrt{\lambda}}
\Big \{ (m_b-m_s)   \Delta
+{(\Delta u - 2 \Delta' s) \over \lambda}
[m_b ( -2 \Delta + 2 \Delta' +u - 2 s') +m_s (2 s -u)] \Big\}
\ee
($\Delta = s-m_b^2$, $\Delta' =s'\, -\,  m_s^2$, $u = s + s' - q^2$ and
$\lambda = u^2-4\,s\,s'$).
The coefficient $d_5$ reads:
\be
d_5 =   {m_b^2 m_s  \over 4 M^{'4} } -{(m_b-m_s) \over 6 M^{'2}}
+{2 m_b  \over 3 M^2 } +{m_b^3  \over 4 M^4 }
+{m_b (m_b^2+m_s^2-m_b m_s -q^2) \over 6 M^2 M^{'2}} \;.
\ee
The integration region $D$ in Eq. (\ref{f0}) is limited by the effective
thresholds $s_0$ and $s_0^\prime$ separating, in the sum rule, the
contribution of the ground state from the hadronic continuum, the latter
being modeled by the leading order perturbative QCD. The parameters employed
in the calculation are the quark masses: $m_b=4.6 \; GeV$ \cite{Dom0}
and $m_s=0.175 \pm 0.020 \; GeV$ \cite{Paver}
(we neglect up and down quark masses),
the leptonic constants $f_B=0.18 \; GeV$
\cite{Dom1}
and $f_K=0.16 \; GeV$. As for the
dimension 3 and dimension 5 condensates, we take
$< {\bar q} q>(1 \; GeV)=(- 230 \; MeV)^3$ and
$<\bar q g_s \sigma^{\mu \nu} G^a_{\mu \nu} {\lambda^a \over 2 } q>
=m_0^2 < {\bar q} q>$, with $m_0^2=0.8 \; GeV^2$; within the final
uncertainties, rescaling
the quark and mixed condensate to higher scales by
the leading-log approximation of the anomalous dimensions
does not affect the numerical results for the
form factors.

To derive $F_0$ from Eq. (\ref{f0}), one looks for a region
where the result is stable under variation of
the Borel parameters
$M^2$, $M^{'2}$, and of the continuum thresholds
$s_0$ for the $B$-channel and  $s^\prime_0$ for the $K$-channel.
We find that stability is obtained
by choosing the continuum thresholds
in the ranges $s_0=33 - 36 \; GeV^2$,
$s^\prime_0=1.3 - 1.5 \; GeV^2$, and the Borel parameters
in the range $M^2=8\pm 1 \; GeV^2$ and
$M^{'2}= 2.0\pm 0.4 \; GeV^2$. Let us finally notice that
the OPE expansion of the correlators in Eqs. (\ref{corrk}) and (\ref{corrks}),
starting from spacelike values of the momentum transferred,
 can be extrapolated to positive $q^2$ provided that
one is far from non-Landau singularities
\cite{Dosch}. We compute
the form factors up to $q^2=15 \; GeV^2$, which is in the safe region.

The result for $F_0(q^2)$
corresponding to the above input numbers is depicted in Fig. 5.

The analogous expressions relevant to $A_0(q^2)$ can be found in
Ref. \cite{Defazio}.
In this case, the same input parameters are used as for $F_0(q^2)$, except
that we need, in this case, $f_{K^*}=0.22 \; GeV$  and the threshold in the
$K^*$-channel is now in the range $s^\prime_0=1.5 - 1.7 \; GeV^2$ (while the
range of $s_0$ for the $B$-channel is the same as before). The results for
$A_0(q^2)$ are displayed in Fig. 6.

As an important virtue of this calculation, we remark that for both
$A_0$ and $F_0$ there exists, in the sum rules, a hierarchical structure in
the OPE expansion, namely the leading perturbative term is numerically
larger than the power corrections. This is a particularly welcomed feature of
the present calculation, from the point of view of this theoretical
framework, in comparison e.g. with the analogous determination of
the form factor $A_2(q^2)$.

Turning to a discussion of the results,  we observe  in $A_0(q^2)$
a sharp increase when $q^2$ varies
in the range $q^2=0 - 15 \; GeV^2$. The form factor
can be fitted by the expression:
\be A(q^2) = {A(0) \over {1 - {q^2 \over M^2_P}}} \label{pole} \ee
\noindent with
$A_0(0)= 0.27 \pm 0.03 $ and $M_P=4.8 \pm 0.2 \; GeV$.
Therefore, the mass of the pole in  $A_0$  is
slightly smaller than the mass of the first singularity in the $q^2$ channel:
$M_{B_s}= 5.375 \pm 0.006 \; GeV$. This behaviour is similar to what
has been observed in Ref. \cite{Defazio}
for the channels $D \to K^*$, $D \to \rho$, and  $B \to \rho$.

On the other hand, the $q^2$-dependence of
$F_0(q^2)$ is rather soft. We obtain
$F_0(0)=0.29 \pm 0.03$, whereas the fitted mass of the pole in Eq. (\ref{pole})
is $M_P\simeq 7.5 \; GeV$, to be compared to the mass of the first resonance
in the $q^2$ channel, a $b {\bar s} (0^+)$  state expected, in
constituent quark models, in the  region near $6 \; GeV$
(in the BWS model the value: $M_{(b {\bar s} )}(0^+)=5.89 \; \mbox{GeV}$
is chosen for the mass of this state).
This behaviour is analogous to that
observed for the form factor
$F_0(q^2)$ in $B \to \pi$, computed
in the framework of the infinite heavy quark mass limit in Ref. \cite{Col}.

These results support
the observation, made in Ref. \cite{Gourd},
 that the assumption of a given $q^2$-dependence
(such as polar, multipolar, etc.) in heavy meson semileptonic form factors
should be more safely confirmed by some explicit theoretical calculation.
In fact, although the dominance of a singularity is reasonable for high
values of $q^2$, it is possible that, for lower $q^2$, the contribution of
higher states or of different dynamical mechanisms modifies the polar
behaviour. The same argument holds for the application of QCD counting rules,
that {\it a priori} are rigorous in the limit of large, spacelike $q^2$.

\subsection*{4. Predictions and conclusions}

In the constituent quark model the leptonic constants of the charmonium system
can be expressed in terms of the $c \bar c$ wave function at the origin
$\Psi(0)$ \cite{Desh}:
\be f_{\eta_c}^2 = 48 {m_c^2 \over M_{\eta_c}^3} |\Psi(0)|^2 \ee
\be f_{J/\Psi}^2 = 12 {1 \over M_{J/\Psi}} |\Psi(0)|^2 \; ;\ee
\noindent
therefore, the ratio $f_{\eta_c}/f_{J/\Psi}$ can be predicted in terms
of the meson masses and of the charm quark mass:
\be {f_{\eta_c} \over f_{J/\Psi}} =
2 m_c \Big( {M_{J/\Psi} \over M_{\eta_c}^3}\Big)^{1\over 2} = 0.97 \pm 0.03 \;,
\label{ratio}\ee
\noindent where the value for $m_c$ chosen in Sec. (2) has been used.
It is known that relativistic and radiative QCD corrections could modify
the prediction in Eq. (\ref{ratio}).
Using our result in
Eq. (\ref{pred}), which includes both, and the experimental value
$f_{J/\Psi}=384 \pm 14 \; MeV$, we get:
\be {f_{\eta_c} \over f_{J/\psi}} = 0.81 \pm 0.05 \; .\ee
\noindent
A comparison of this result with that in Eq. (\ref{ratio})
suggests that corrections are at the level of $15- 20 \%$.
On the other hand, Eq. (\ref{ratio}), applied to the radial
excitations $\eta^\prime_c$ and $\Psi^\prime$, gives the prediction:
${f_{\eta^\prime_c} \over f_{\Psi^\prime}} = 0.80 \pm 0.02$, to be compared to
${f_{\eta^\prime_c} \over f_{\Psi^\prime}} = 0.87 \pm 0.08$,
obtained from Eq. (\ref{pred}) and using the experimental measurement
$f_{\Psi^\prime}=282 \pm 14 \; MeV$,
possibly suggesting a minor role of radiative and relativistic corrections for
these states.

Turning to nonleptonic $B$ decays, we first analyze
 ratios of
decay widths, such as  $B \to K^{(*)} \eta_c$ and
 $B \to K^{(*)} \eta_c^\prime$, in the
factorization approximation, where the dependence on the Wilson
coefficient $a_2$ and on other weak parameters drops out,
the relevant remaining dynamical quantities being the leptonic
constants and the semileptonic form factors.

Consider, for example, the ratio:
\be {\tilde R}_K = { \Gamma(B^- \to K^- \eta^\prime_c) \over
\Gamma(B^- \to K^- \eta_c)} = \; 0.771 \;
\big({f_{\eta^\prime_c} \over f_{\eta_c}}\big)^2 \;
\big({F_0(M^2_{\eta^\prime_c}) \over
F_0(M^2_{\eta_c})} \big)^2 \ee
\noindent where the numerical term is a phase space factor.
The interesting point is that, because of the flat shape of
$F_0(q^2)$, ${\tilde R}_K$ mainly depends
on the ratio of the leptonic constants:
\be {\tilde R}_K = 0.771 \; ({f_{\eta^\prime_c} \over f_{\eta_c}}\big)^2
\; (1.09 \pm 0.09)^2 = 0.60 \pm 0.15  \; . \ee
\noindent
Thus, in the factorization approximation, a measurement of ${\tilde R}_K$
would provide us with interesting information on
${f_{\eta^\prime_c} \over f_{\eta_c}}$, and complement our knowledge of the
properties of the ${q\bar q}$ wavefunction.

On the other hand, the analogous ratio for the decays into $K^*$ is given by
\be {\tilde R}_{K^*} = { \Gamma(B^- \to K^{*-} \eta^\prime_c) \over
\Gamma(B^- \to K^{*-} \eta_c)} =
0.381 \; \big({f_{\eta^\prime_c} \over f_{\eta_c}}\big)^2 \;
\big({A_0(M^2_{\eta^\prime_c})
\over A_0(M^2_{\eta_c})} \big)^2 =
0.381 \; \big({f_{\eta^\prime_c} \over f_{\eta_c}}\big)^2 \; (1.4 \pm 0.2)^2
\; .\label{polar}\ee
\noindent
Here, the ratio of the form factors deviates from unity due to the
$q^2$-dependence of $A_0$. The prediction from (\ref{polar})
would be:
${\tilde R}_{K^*} = 0.45 \pm 0.16$. Moreover, we observe that in the assumed
factorization approximation
the quantity $\sqrt{ {\tilde R}_{K^*}/{\tilde R}_{K}}$
is sensitive to the $q^2$-dependence of the ratio $A_0/F_0$:
\be
1.42 \; \sqrt{  { {\tilde R}_{K^*}\over {\tilde R}_{K} } }=
\Big( { A_0(M^2_{\eta^\prime_c})/F_0(M^2_{\eta^\prime_c}) \over
A_0(M^2_{\eta_c}) /F_0(M^2_{\eta_c}) } \Big)
\ee
\noindent i.e. mainly to the $q^2$-dependence of $A_0$ since $F_0$ is
predicted to have a rather flat behaviour.

A bound on the ratio
\be
{R}_{\eta_c} = { \Gamma(B^- \to K^{*-} \eta_c) \over
\Gamma(B^- \to K^- \eta_c)}= 0.373 \;
\big({A_0(M^2_{\eta_c})
\over F_0(M^2_{\eta_c})} \big)^2
\ee
\noindent has been derived in Ref. \cite{Gourdin1}:
$0.19 \le { R}_{\eta_c} \le 0.98$. This bound is satisfied by our result
${R}_{\eta_c}= 0.73 \pm 0.13$. In addition,
 for the analogous  quantity ${ R}_{\eta^\prime_c}$, we predict:
${ R}_{\eta^\prime_c}=0.56 \pm 0.12$.

Finally, we consider the ratio:
\be R_K= { \Gamma(B^- \to K^{-} \eta_c) \over
\Gamma(B^- \to K^{-} J/\Psi)} =
2.519 \; \big({f_{\eta_c} \over f_{J/\Psi}}\big)^2 \;
\big({F_0(M^2_{\eta_c})
\over F_1(M^2_{J/\Psi})} \big)^2 \;.
\ee
\noindent In order to predict $R_K$ we can use the relation:
$F_1(0) = F_0(0)$, and the observation, common to the
QCD sum rules calculations
of the $q^2$-dependence of $F_1$, of the validity of the single pole
model dominated by the $1^-$ $B^*_s$ resonance
\cite{Dosch, Col, Ball}. We obtain:
$R_K=0.94 \pm 0.25$, and, for the analogous quantity
$R^\prime_K= { \Gamma(B^- \to K^{-} \eta^\prime_c) \over
\Gamma(B^- \to K^{-} \Psi^\prime)}$:
$R^\prime_K= 1.61 \pm 0.53$.
This implies that, using the CLEOII experimental measurements:
${\cal B}(B^- \to K^{-} J/\Psi) = (0.11 \pm 0.01 \pm 0.01)\times 10^{-2}$
and
${\cal B}(B^- \to K^{-} \Psi^\prime) = (0.06 \pm 0.02 \pm 0.01)\times 10^{-2}$
we expect:
${\cal B}(B^- \to K^{-} \eta_c) = (0.11 \pm 0.03)\times 10^{-2}$
and
${\cal B}(B^- \to K^{-} \eta_c^\prime) = (0.10 \pm 0.05)\times 10^{-2}$,
and therefore these decays are in a
range well-accessible to the present experimental facilities.

With these predictions we conclude our analysis. The relatively large decay
rates of the processes $B \to K^{(*)} \eta_c$ and
$B \to K^{(*)} \eta_c^\prime$  should allow their observation
in the near future.
This measurement will shed more light on the problem of factorization,
which is a basic assumption in the present analysis of heavy meson nonleptonic
decays.

{}From the theoretical point of view, it should be interesting to compare the
results obtained here with other
QCD calculations. For example, it should be possible to
calculate the leptonic constants of $\eta_c$ and $\eta^\prime_c$ mesons by
lattice QCD. Moreover, regarding the $B \to K^{(*)}$ form factors, a possible
independent test of the $q^2$-dependence could be obtained, e.g., by using
light-cone sum rules. Also, extrapolations of form factors in $q^2$ and in
the heavy quark mass, starting from the charm mass, in the lattice QCD
framework \cite{lattice},
could possibly provide us with enough information,
at least at a qualitative level,
to be compared with the QCD sum rules results.

\newpage
\noindent {\bf Acknowledgments \\}
We would like to thank F.Buccella, F.De Fazio,
G.Nardulli and P.Santorelli for interesting discussions.
The work of (NP) was supported in part by the HCM, EEC Contract
ERBCHRXCT930132.
The work of (CAD) was supported in part by the
Foundation for Research Development (ZA) and the John Simon Guggenheim
Memorial Foundation (USA).

\newpage

\hskip 3 cm {\bf FIGURE CAPTIONS}
\vskip 1 cm
{\bf Fig. 1} \par
\noindent
The mass of the $\eta_{c}$ from Hilbert moments as a function of $n$
for $\Lambda = 200 \; \mbox{MeV}$, $m_{c} = 1.39 \; \mbox{GeV}$, and
$<\alpha_{s} G^{2}> = 0.063 \; \mbox{GeV}^{4}$. The solid curve is the
prediction and the dash curve the experimental value.
\vskip 1 cm
{\bf Fig. 2} \par
\noindent
The leptonic decay constant
${\tilde f}_{\eta_{c}}=f_{\eta_{c}}/\sqrt 2$,
using a single pole spectral
function in Hilbert moments, as  a function of $n$ and for the same
values of input parameters as in Fig. 1. The solid (dash) curve is obtained
using the predicted (experimental) value of $M_{\eta_{c}}$.
\vskip 1 cm
{\bf Fig. 3} \par
\noindent
The mass of the $\eta_{c}$ from the Laplace transform
for $\Lambda = 200 \; \mbox{MeV}$, $m_{c} = 1.39 \; \mbox{GeV}$, and
$<\alpha_{s} G^{2}> = 0.063 \; \mbox{GeV}^{4}$. The solid curve is the
prediction and the dash curve the experimental value.
\vskip 1 cm
{\bf Fig. 4} \par
\noindent
The leptonic decay constant
${\tilde f}_{\eta_{c}}=f_{\eta_{c}}/\sqrt 2$,
using a single pole spectral
function in the Laplace transform, and for the same
values of input parameters as in Fig. 3. The solid (dash) curve is obtained
using the predicted (experimental) value of $M_{\eta_{c}}$.
\vskip 1 cm
{\bf Fig. 5} \par
\noindent
The form factor $F_0(q^2)$ for the transition $B \to K$.
The curves refer to the sets of parameters:
$s_0=33 \; GeV^2$ and $s^\prime_0=1.3 \; GeV^2$ (continuous line),
$s_0=33 \; GeV^2$ and $s^\prime_0=1.5 \; GeV^2$ (dashed line),
$s_0=36 \; GeV^2$ and $s^\prime_0=1.3 \; GeV^2$ (dotted line),
$s_0=36 \; GeV^2$ and $s^\prime_0=1.5 \; GeV^2$ (dashed-dotted line).
The Borel parameters in the $B$ and $K$ channel are fixed to
$M^2=8 \; GeV^2$, $M^{\prime 2}=2  \; GeV^2$, respectively.
\vskip 1 cm
\vskip 1 cm
{\bf Fig. 6} \par
\noindent
The form factor $A_0(q^2)$ for the transition $B \to K^*$.
The curves refer to the sets of parameters:
$s_0=33 \; GeV^2$ and $s^\prime_0=1.5 \; GeV^2$ (continuous line),
$s_0=33 \; GeV^2$ and $s^\prime_0=1.7 \; GeV^2$ (dashed line),
$s_0=36 \; GeV^2$ and $s^\prime_0=1.5 \; GeV^2$ (dotted line),
$s_0=36 \; GeV^2$ and $s^\prime_0=1.7 \; GeV^2$ (dashed-dotted line).
The Borel parameters in the $B$ and $K^*$ channel are
the same as in Fig.5.

\end{document}